\documentclass[submission,copyright,creativecommons]{eptcs}
\providecommand{\event}{ACL2 Workshop 2018} 
\usepackage{underscore}           
\usepackage{breakurl}             
\usepackage{amsmath,amssymb,amsfonts,amsthm}
\usepackage{fancyvrb}
\usepackage[shortcuts]{extdash}
\usepackage{color}
\usepackage{framed}
\usepackage{xspace}
\usepackage{listings}
\title{DefunT: A Tool for Automating Termination Proofs by Using the Community Books (Extended Abstract)}
\author{Matt Kaufmann
\institute{Department of Computer Science\\
The University of Texas at Austin, Austin, TX, USA}
\email{kaufmann@cs.utexas.edu}}
\def\titlerunning{DefunT}
\def\authorrunning{M. Kaufmann}
\begin{document}
\maketitle

\begin{abstract}

We present a tool that automates termination proofs for recursive
definitions by mining existing termination theorems.

\end{abstract}

The macro, {\tt defunT} ({\tt \bf defun} with
auto-{\tt\bf{T}}ermination), is a tool that can automate ACL2 proofs
of termination (i.e., of measure conjectures).  This note has three
goals: to introduce this tool to potential users, to explain some of
its implementation, and to advertise for research collaborators to
improve the tool.  The tool suite resides in community books directory
\verb|kestrel/auto-termination/|.\footnote{An archival version, from
  the time this paper was written, is under
{\tt books/workshops/2018/kaufmann/}.}

{\tt DefunT} relies on a database of already-proved termination theorems, each stored
as a list of clauses (disjunctions).  That database is generated by
the script file \verb|write-td-cands.sh|, which writes it to the
generated book, \verb|td-cands.lisp|, and creates an associated file,
\verb|td-cands.acl2|.  This script computes the database after it
includes the book \verb|doc/top.lisp|, which in turn includes many of
the community books (to build the manual), using algorithms
implemented in the book, \verb|termination-database.lisp|.  The book
\verb|td-cands.lisp| will likely only need to be regenerated
infrequently; but it is routinely certified by the build system on top
of a world obtained by executing the 45 {\tt include-book} events in
\verb|td-cands.acl2|, which define all necessary packages so that ACL2
can read all forms in the book.

We explain {\tt defunT} --- both its use and a little about its
implementation --- by focusing on the following example, which creates
three distinct proof goals for termination: one for each recursive
call.  The book \verb|defunt-top.lisp| includes both the database,
\verb|td-cands.lisp|, and the implementation of the {\tt defunT} macro,
\verb|defunt.lisp|.

{\footnotesize\begin{verbatim}
(include-book "kestrel/auto-termination/defunt-top" :dir :system)
(defunt f3 (x y)
  (if (consp x)
      (if (atom y)
          (list (f3 (cddr x) y) (f3 (cadr x) y))
        (f3 (cdr x) y))
    (list x y)))
\end{verbatim}}

\noindent The output shown below notes that {\tt defunT} finds three
helpful termination theorems in the database, \verb|td-cands.lisp|.
Each of these suffices to prove one of the three goals with a
{\tt :termination-theorem} lemma-instance, where one of those three
requires a book to be included.

{\footnotesize\begin{verbatim}
     *Defunt note*: Using termination theorems for SYMBOL-BTREE-TO-ALIST-AUX,
     EVENS and TRUE-LISTP.

     *Defunt note*: Evaluating 
     (LOCAL (INCLUDE-BOOK "misc/symbol-btree" :DIR :SYSTEM))
     to define function SYMBOL-BTREE-TO-ALIST-AUX.
\end{verbatim}}

\noindent The {\tt defunT} macro uses {\tt make-event} to do the
search and to generate a suitable event as displayed below.  The
search can make two passes through the database, where the first pass
only considers functions defined in the current session.  In this
example, a local {\tt include-book} is generated because the first
pass was not sufficient.  In spite of making both passes, ACL2 reports
only 0.04 seconds taken altogether, using a 2014 MacBook Pro.

{\footnotesize\begin{verbatim}
(ENCAPSULATE ()
  ;; The following book is necessary, as noted in the output shown above.
  (LOCAL (INCLUDE-BOOK "misc/symbol-btree" :DIR :SYSTEM))
  ;; Six local defthm events are omitted here.  The seventh has the following form:
  (LOCAL (DEFTHM new-termination-theorem
           <termination theorem for f3 generated from found measure etc.>
           :HINTS (("Goal" :USE <..elided here..> :IN-THEORY (THEORY 'AUTO-TERMINATION-FNS)))))
  (DEFUN F3 (X Y)
    (DECLARE (XARGS :MEASURE (ACL2-COUNT X)
                    :HINTS (("Goal" :BY (:FUNCTIONAL-INSTANCE new-termination-theorem
                                                              (binary-stub-function F3))))))
    (IF (CONSP X)
        (IF (ATOM Y)
            (LIST (F3 (CDDR X) Y) (F3 (CADR X) Y))
            (F3 (CDR X) Y))
        (LIST X Y))))
\end{verbatim}}

\noindent A key aspect of {\tt defunT} is that termination theorem
clause-lists are stored in {\em simplified} form: thus, an old
clause-list can subsume a new clause even when function bodies have
minor differences, such as {\tt (if (endp x) ...)} vs. {\tt (if (not
  (consp x)) ...)}.  Also, the generated local theorems are carefully
instrumented to make proofs fast and automatic.  The flow is as
follows (here, restricting to the case of a single old termination
theorem), where $old$ and $new$ are old and new termination theorems,
and $old_s$ and $new_s$ are their simplifications: $new$ follows with
a {\tt :use} hint from $new_s$, which follows with a {\tt :by} hint
from $old_s$, which follows with a {\tt :use} hint from $old$.  The
{\tt :by} hint has two advantages over a corresponding {\tt :use}
hint: it avoids the need to supply a substitution (when the old and
new functions have different formals), and it avoids if-splitting into
clauses (goals).  The {\tt :by} hint succeeds because it employs
essentially the same subsumption test as is used during the search for
an old termination theorem to prove the new termination goal.  The
{\tt :use} hints are accompanied by {\tt :in-theory} hints that can be
expected to make those proofs fast, by restricting to the small
theories used for clause-list simplification.  Stub functions replace
functions called in their own termination schemes, to enhance
subsumption.

{\em Conclusion.}  Program termination is a rich
field~\cite{Cook:2011:PPT:1941487.1941509}.  The goal of {\tt defunT}
is, however, simply to make it convenient to prove termination
automatically when using ACL2.  An extension of ACL2 with CCG
analysis~\cite{ccg} can prove termination automatically; unlike that
approach, {\tt defunT} generates a measure for ACL2's usual
termination analysis.  J Moore's tool
Terminatricks~\cite{terminatricks} is a different step towards that
goal: while that tool does not use the {\tt defunT} approach of taking
advantage of the community books, it can however incrementally extend
its database of termination theorems.  This potential enhancement to
{\tt defunT} is discussed in file {\tt to-do.txt}, as are more than 20
others.  Further implementation-level details may be found in the
\verb|README|, which for example explains database organization by
     {\em justification} (which includes a measure), as well as
     several optimizations, such as the use of subsumption to restrict
     the database to 643 distinct termination schemes essentially
     shared by 821 functions.  Others are invited to contribute to the
     enhancement of {\tt defunT}!

{\em Acknowledgments.}  This material is based upon work supported in
part by DARPA under Contract No. FA8750-15-C-0007.  We thank Eric
Smith for encouragement to work on this problem.

\nocite{*}
\bibliographystyle{eptcs}
\bibliography{defunt-kaufmann}

\begin{thebibliography}{1}
\providecommand{\bibitemdeclare}[2]{}
\providecommand{\surnamestart}{}
\providecommand{\surnameend}{}
\providecommand{\urlprefix}{Available at }
\providecommand{\url}[1]{\texttt{#1}}
\providecommand{\href}[2]{\texttt{#2}}
\providecommand{\urlalt}[2]{\href{#1}{#2}}
\providecommand{\doi}[1]{doi:\urlalt{http://dx.doi.org/#1}{#1}}
\providecommand{\bibinfo}[2]{#2}

\bibitemdeclare{article}{Cook:2011:PPT:1941487.1941509}
\bibitem{Cook:2011:PPT:1941487.1941509}
\bibinfo{author}{Byron \surnamestart Cook\surnameend}, \bibinfo{author}{Andreas
  \surnamestart Podelski\surnameend} \& \bibinfo{author}{Andrey \surnamestart
  Rybalchenko\surnameend} (\bibinfo{year}{2011}): \emph{\bibinfo{title}{Proving
  Program Termination}}.
\newblock {\sl \bibinfo{journal}{Commun. ACM}}
  \bibinfo{volume}{54}(\bibinfo{number}{5}), pp. \bibinfo{pages}{88--98}.
\newblock \urlprefix\url{http://doi.acm.org/10.1145/1941487.1941509}.

\bibitemdeclare{inproceedings}{ccg}
\bibitem{ccg}
\bibinfo{author}{Matt \surnamestart Kaufmann\surnameend},
  \bibinfo{author}{Panagiotis \surnamestart Manolios\surnameend},
  \bibinfo{author}{J~Strother \surnamestart Moore\surnameend} \&
  \bibinfo{author}{Daron \surnamestart Vroon\surnameend}
  (\bibinfo{year}{2006}): \emph{\bibinfo{title}{Integrating CCG analysis into
  ACL2}}.
\newblock In: {\sl \bibinfo{booktitle}{Workshop Proceedings: WST 2006, Eighth
  International Workshop on Termination}}, pp. \bibinfo{pages}{64--68}.
\newblock
  \urlprefix\url{http://citeseerx.ist.psu.edu/viewdoc/download?doi=10.1.1.97.8994&rep=rep1&type=pdf}.

\bibitemdeclare{misc}{terminatricks}
\bibitem{terminatricks}
\bibinfo{author}{{J} \surnamestart Moore\surnameend} (\bibinfo{year}{Accessed:
  2018}): \emph{\bibinfo{title}{Terminatricks}}.
\newblock
  \bibinfo{howpublished}{{\url{https://github.com/acl2/acl2/tree/master/books/projects/codewalker/terminatricks.lisp}}}.

\end{thebibliography}
\end{document}